# On generalized models of earthquake evolution by the example of classical and mirror triads

A.V. Guglielmi, and O.D. Zotov

*Institute of Physics of the Earth, RAS, Moscow, Russia*

*guglielmi@mail.ru, ozotov@inbox.ru*

**Abstract**

The analysis of the classical and mirror triads of the sequence of earthquakes has been carried out in order to find the equations of evolution of foreshocks and aftershocks. The differential equation with cubic (quadratic) nonlinearity has been proposed to describe the activity of foreshocks (aftershocks). The solution of the governing equation for foreshocks indicates the explosive instability of rocks, leading to the main rupture and the main shock of the earthquake. It is suggested that explosive instability leads to a first-order phase transition, in which the main parameters of the earthquake source as a thermodynamic system change abruptly. A direction for further research has been outlined.

*Keywords*: Omori law, evolution equation, earthquake triads, foreshocks, main shock, aftershocks, rise and fall in seismic activity.

## 1. Introduction

Without claiming to be philosophical revelations, we simply recall the idealistic point of view, according to which the laws of nature are based on certain mathematical relationships. It is appropriate to immediately give an example related to the topic of this paper. The first quantitative law of earthquake physics (Omori's law) states:

$$n = \frac{k}{c+t}. \qquad (1)$$

Here $n(t)$ is the frequency of aftershocks, $t$ is time, $k > 0$, $c > 0$, $t \geq 0$ [1]. Reviews [2–4] contain detailed information about the outstanding discovery of Fusakichi Omori made in 1894.

Mathematically, one can express Omori's law in the form of a differential equation

$$\frac{dn}{dt} + \sigma n^2 = 0, \qquad (2)$$

where $\sigma = 1/k$ is the deactivation coefficient of the earthquake source [5]. Law (2) looks simpler, since it uses three symbols $n$, $t$, $\sigma$, and not four, as in the original Omori formula (1). The equation is non-linear, but we can make a linearization. Let us change the independent variable $n \to g = 1/n$ and write the aftershock equation in the simplest form of a linear equation

$$\dot{g} = \sigma. \qquad (3)$$

Thus we have three expressions for the same law. Formally, all three forms of the law are equivalent to each other, but the axiological function of the evolution equations (2), (3) attracted our special attention, and together with colleagues, we published a series of studies based on the aftershock evolution equation [6–14].

Here it is appropriate to note that any law always limits arbitrariness in some sense. It was not for nothing that we presented the simplest form (3) of Omori's law. After all, the law in its original form (1) unambiguously prescribes a strictly linear growth of the variable $g$ over time. The entry of the law in the form (3) indicates that this is too severe a limitation. However, in our studies, we only gradually, step by step, removed the restrictions, and first of all assumed that the deactivation coefficient $\sigma$ can be a function of time. Then instead of the Omori formula (1) we have a more flexible formula

$$n(t) = n_0 \left[ 1 + n_0 \int_0^t \sigma(t') dt' \right]^{-1}, \qquad (4)$$

in which $n_0 = n(0)$.

The next step was to add the linear term $-\gamma n$ to the left side of equation (2). As a result, we obtain the Verhulst logistic equation, which takes into account an important feature of the aftershock flow [10]. If we now add term $-\kappa \partial^2 n(t,x)/\partial x^2$, then we will arrive at the classical Kolmogorov-Petrovsky-Piskunov equation [11, 14]. This makes it possible to understand the aftershock migration discovered in [7]. By adding a Langevin source to the right side of (2), we arrive at a stochastic evolution equation describing fluctuations in the aftershock flux.

In this paper, an attempt is made to describe mathematically the evolution of foreshocks in a similar way.

## 2. Curves of growth and decline in activity

Foreshocks, the main shock of an earthquake, and aftershocks form a unique unity called a triad in [15]. The magnitude of the main shock always exceeds the magnitudes of foreshocks and aftershocks. In the classical triad, the number of aftershocks is greater than the number of foreshocks. Sometimes foreshocks are not observed at all. In this case, it is natural to call the triad a shortened (truncated) classical one.

A rare but extremely interesting variety of mirror triads has recently been discovered [16]. In the mirror triad, the number of foreshocks significantly exceeds the number of aftershocks. We use shortened mirror and classical triads to analyze the rise and fall of tremor activity.

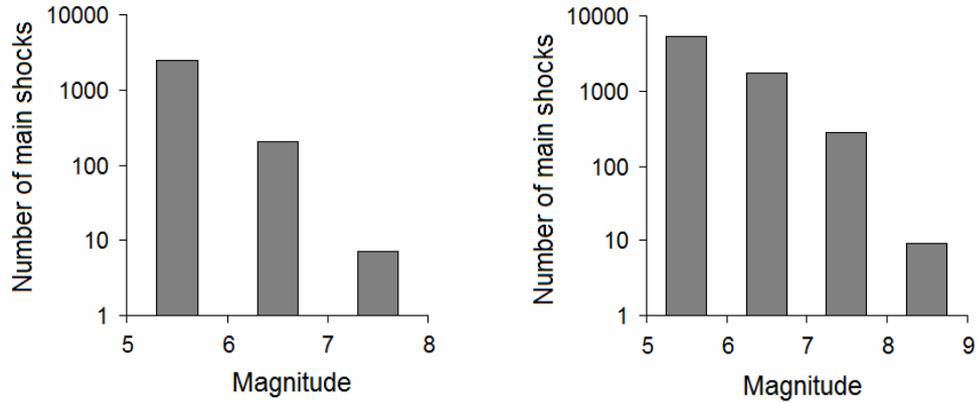

Fig. 1. Distribution over the magnitude of the main shocks for shortened mirror (left) and classic (right) triads.

Two distributions over the magnitude of the main strokes are shown in Figure 1. On the left panel (right panel), distribution is given for all cases of the appearance of a shortened mirror triad (classic triad). We used here and below the data on earthquakes that occurred on Earth from 1973 to 2019, and were presented in the USGS/NEIC world catalog of earthquakes (https://earthquake.usgs.gov). The main shocks with a magnitude $M \geq 5$ and a hypocenter depth not exceeding 250 km were selected. For each main shock there was a circular epicentral zone was determined according to the method described in [17]. In total, 2682 main shockes for shortened mirror triads and 7345 for classic shortened triads have been accumulated.

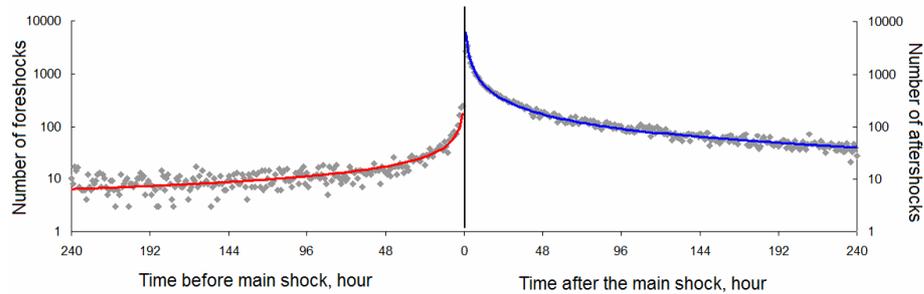

Fig. 2. Generalized evolution of foreshocks and aftershocks in shortened triads. The zero point corresponds to the moments of the main shocks. To the left and to the right of the zero point, two semi-axes of time depart. One is directed to the past, the other to the future.

Figure 2 shows a generalized picture of the growth and decline in the activity of earthquakes before and after the main shock, respectively. The figure was built using the epoch superposition method. The moments of the main shocks are chosen as the reference point. Time in the figure is represented by two semi-axes, one of which is directed to the past, and the other to the future. The dots in the figure indicate the number of events in the hourly interval. The total number of foreshocks (aftershocks) is 3903 (43001). Curves $n(t) = at^{-b}$ approximate the experimental points.

For foreshocks (aftershocks) $a = 178$ ($a = 6228$), $b = 0.6$ ($b = 0.92$), and the coefficient of determination is equal to $R^2 = 0.74$ ($R^2 = 0.97$).

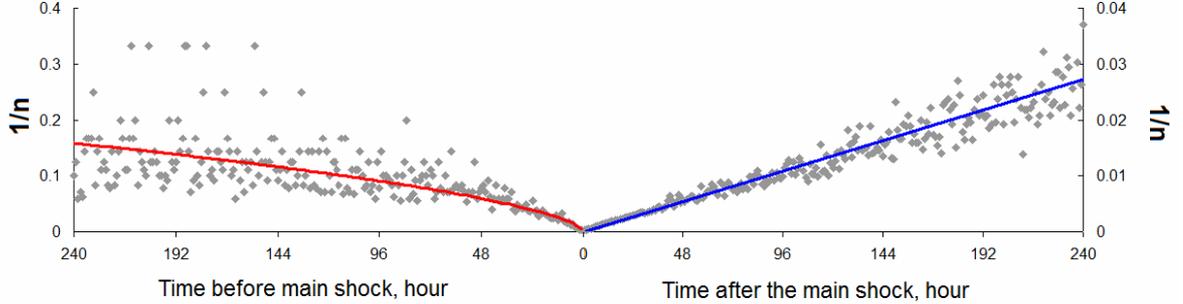

Fig. 3. Auxiliary functions $g(t) = 1/n(t)$ for foreshocks and aftershocks in shortened triads. The zero point corresponds to the moments of the main shocks. To the left and to the right of the zero point, two semi-axes of time depart. One semi-axis is directed to the past, the other to the future.

The time dependence of the auxiliary function $g(t)$ clearly shows a nontrivial distinction between the evolution of foreshocks and aftershocks (see Figure 3). We clearly see that foreshock activity increases in a completely different way than aftershock activity decreases. The approximation for foreshocks is $g = 0.0056 t^{0.6}$ ($R^2 = 0.745$), while for aftershocks $g = (t-2)10^{-4}$ ($R^2 = 0.74$).

## 3. Discussion

The aftershock evolution equation (3) predicts a linear increase in $g$ over time, which corresponds to the Omori hyperbolic law (1). The right branch in Figure 3 unconditionally confirms the correctness of equation (3) and, thus, the correctness of equation (2), which contains a quadratic non-linearity.

The left branch in Figure 3 clearly indicates that in the case of foreshocks, we are dealing with a higher-order nonlinearity. The simplest hypothesis is that the master equation for foreshocks has the form of a first-order differential equation with cubic nonlinearity:

$$\frac{dn}{dt} + \tau n^3 = 0. \qquad (5)$$

When writing (5), we made the substitution $t \to -t$, i.e. the semi-axis of time in (5), as in Figure 3, is directed to the past. The phenomenological parameter $\tau$ has the dimension of time.

The inverse problem of the physics of the earthquake source is to estimate the phenomenological parameters from the observational data of earth tremors [14]. When estimating the deactivation coefficient $\sigma$, regularization is preliminarily performed, which consists in smoothing the initial data $g \to \langle g \rangle$. The dimensionless quantity $\sigma$ is determined by the formula

$$\sigma = \frac{d}{dt}\langle g \rangle. \tag{6}$$

The parameter $\tau$ can be estimated in a similar way. The corresponding formula has the form

$$\tau = \frac{1}{2}\frac{d}{dt}\langle g^2 \rangle. \tag{7}$$

The time inversion operation $t \to -t$ is formally allowed, but physically meaningless. Let's return to real time and rewrite the evolution equation (5):

$$\frac{dn}{dt} - \tau n^3 = 0. \tag{8}$$

The solution of the equation has the form

$$n = \frac{1}{\sqrt{-2\tau t}}, \tag{9}$$

or $g(t) = g_0 \sqrt{t/t_0}$ where $g_0 = g(t_0)$, $t_0 = -1/2\tau n_0^2$, $n_0 = n(t_0)$, $t < 0$. We see that $n \to \infty$ when $t \to 0_-$. In other words, our theory points to the explosive instability of the rocks, leading to the main rupture and the main shock of the earthquake. (Recall that an explosive instability is called an instability in which the singularity is reached in a finite time interval. For explosive and exponential non-stability in geophysical media, see [10, 18, 19].)

Apparently, at $t = 0$, the first-order phase transition occurs. Indeed, judging by Figure 2, there is a jump in the frequency of tremors when passing through the zero reference point: $\Delta n = n(0_+) - n(0_-)$. There is no doubt in our minds that the jump in $n$ indicates an abrupt change in the main parameters of the source as a thermodynamic system. These assumptions (explosive instability and phase transition) deserve further study. In particular, we expect that in the so-called symmetric triads [16] a second-order phase transition will be observed. The so-called GTS (*Grande terremoto solitario*) [14] are also interesting in this respect. The phase transition probably occurs in this case as well, but the main shock does not manifest itself in any way in tremors either before or after GTS.

Concluding this section of the paper, we want to say that from a mathematical point of view, our evolution equations (2) and (8) are quite simple. And yet they give us an interesting example of mathematical minimalism in describing earthquakes. Much more important, however, is that our theory is open to falsification and generalization. Let us add linear terms $\gamma_\pm n$ to (2), (8) and write the control equation of the source in the following form as the evolution equation of a dynamic system:

$$\frac{dn}{dt} = -\frac{\partial U}{\partial n}. \tag{10}$$

The potential $U(n; \sigma, \tau, \gamma_\pm)$ is $U_- = \gamma_- n^2/2 - \tau n^4/4$ for $t < 0$, and $U_+ = -\gamma_+ n^2/2 + \sigma n^3/3$ for $t > 0$.

## 4. Conclusion

The choice of the direction of this study was motivated by a natural desire to imagine, at least in general terms, what the foreshock evolution equation might look like. Due to the small number of foreshocks in the classical and even in the mirror triad, we used the superposed epoch analysis in order to accumulate a sufficiently large amount of initial data for statistical study. As a result, we, firstly, confirmed the acceptability of the evolution equation (2), which contains a quadratic nonlinearity. Second, we found one characteristic difference between the nonlinear evolution of foreshocks and aftershocks. This observation made it possible to hypothesize that the master equation for foreshocks has the form of the differential equation (5) with cubic nonlinearity. The solution of the equation testifies in favor of the hypothesis of a phase transition, to which the explosive instability of rocks in the earthquake source leads. We get the impression of the contours of a single entity, which includes the transition from the past (foreshock epoch) to the future (aftershock epoch, or Omori epoch, as we called it in due time [8]).

In this paper, we presented the analysis of shortened triads. Further research will be directed to the analysis of complete triads, mirror and classical. An extended version of this paper will be submitted to the Journal of Volcanology and Seismology [20].

*Acknowledgments*. We express our sincere gratitude to A.D. Zavyalov and B.I. Klain for support and fruitful discussions.